**To be Published in Applied Physics Letter Soon.**

# SPIN TRANSFER SWITCHING AND SPIN POLARIZATION IN MAGNETIC TUNNEL JUNCTIONS WITH MgO AND $AlO_x$ BARRIERS

Zhitao Diao, Dmytro Apalkov, Mahendra Pakala, Alex Panchula, and Yiming Huai

Grandis Inc., 1266 Cadillac Court, Milpitas CA 95035

We present spin transfer switching results for MgO based magnetic tunneling junctions (MTJs) with large tunneling magnetoresistance (TMR) ratio of up to 150 % and low intrinsic switching current density of 2-3 x $10^6$ A/cm$^2$. The switching data are compared to those obtained on similar MTJ nanostructures with $AlO_x$ barrier. It is observed that the switching current density for MgO based MTJs is 3-4 times smaller than that for $AlO_x$ based MTJs, and that can be attributed to higher tunneling spin polarization (TSP) in MgO based MTJs. In addition, we report a qualitative study of TSP for a set of samples, ranging from 0.22 for $AlO_x$ to 0.46 for MgO based MTJs, and that shows the TSP (at finite bias) responsible for the current-driven magnetization switching is suppressed as compared to zero-bias tunneling spin polarization determined from TMR.

The advantages of the next generation magnetic random access memory (MRAM) and the high frequency oscillators that utilize spin-transfer phenomenon have led to a flurry research activity in recent years [1-5]. Spin transfer switching in magnetic tunneling junctions (MTJs) is of particular interest because of large read signal and large cell resistance typically observed in MTJs [2-4]. Achieving low intrinsic switching current density ($J_{c0,}$ ~ $10^5$ A/cm$^2$) in MTJs is the most challenging issue for successful adaptation of the spin-transfer switching as a writing scheme in high–density MRAM. MTJs using $AlO_x$ barriers combined with dual spin filter structures and low saturation moment free layers such as CoFeB [3] have been investigated for significant $J_{c0}$ reduction. Spin transfer switching study in recently reported MgO MTJs with large TMR [4] is of increasing interest because lower intrinsic current density $J_{c0}$ is expected due to much higher spin polarization founded in the MgO based MTJs compared to AlOx based MTJs [3]. This letter reports the spin transfer switching results at room temperature (RT) on low resistance area product (RA) MgO MTJ nanostructures, along with those obtained from $AlO_x$ MTJ ones. The MTJ samples display a room temperature (RT) TMR ranging from 15% to 150% and switching current density of 2-10 MA/cm$^2$ with low RA ranging from 10 to 50 Ω·μm$^2$. All samples in this study used CoFeB free layers with a low saturation moment of 1050 emu/cm$^3$. Specifically, this investigation examines the TMR bias dependence and tunnel spin-polarization (TSP) correlation to the current-driven spin transfer effect and intrinsic switching current density $J_{c0}$.

Magnetic tunneling junction films used in this study were of the form Ta5/PtMn30/CoFe3/ Ru0.8/CoFeB2/$AlO_x$/CoFeB2.5/Ta5 (nm) and Ta5/PtMn20/CoFe2 /Ru 0.8/CoFeB 2/MgO /CoFeB 2.5/ Ta 8 (nm). They were deposited in a Singulus PVD cluster system and annealed at 250-350$^o$C for 1.5-2 hours in a magnetic field of 1 Tesla. The MTJ films were subsequently patterned into deep submicrometer ellipse shaped pillars in methods described elsewhere [2, 6]. A quasistatic tester with pulse current capability was used to measure resistance ($R$) as a function of magnetic field ($H$) and current ($I$) at room temperature. The $R$ vs. $I$ data for current switching measurement were obtained in pulse mode by sequentially increasing/decreasing the amplitude of current pulse in steps of 50 μA for different current pulse widths between 3 msec and 1 sec. The MTJ resistance value was measured after each pulse step using a low read current of 10 μA, which does not affect the magnetic state of the sample. The offset fields $H_{off}$, experienced by the free layers due to the orange – peel coupling field and the dipolar field from the adjacent pinned layers, were balanced by applying an external field $H_a=H_{off}$ during switching current measurements [3]. The bias dependence of resistance (or TMR) was measured in a DC mode with a current increment in step of 100 μA.

A typical field hysteresis loop ($R$ vs. $H$) and current loop ($R$ vs. $I$) for an MTJ cell with an MgO barrier of 1.2 nm are shown in Fig. 1a and 1b. The nominal magnetic cell dimension is 125 nm x 205 nm. TMR of ~150% is calculated from the $R$ vs. $H$ plot. Average RT switching current ($\langle I_c \rangle$) is 0.22 mA at a pulse width of 30 ms, where $I_c$ is defined as $\left[I_c^{P\to AP} - I_c^{AP\to P}\right]/2$, and $I_c^{P\to AP(AP\to P)}$ is the current required to switch free layer magnetization from the parallel (anti-parallel) to anti-parallel (parallel) state. The RT current density ($J_c$) was calculated from $\langle I_c \rangle$ to be 1.1 MA/cm$^2$ and RA is ~50 Ω·μm$^2$. These results can be compared to switching data for $AlO_x$ based MTJ cell. A typical field hysteresis loop and current loop for an MTJ cell with an $AlO_x$ barrier of 0.5 nm are shown in Fig. 1c and 1d. The nominal magnetic cell dimension is 130 nm x 170 nm and a TMR of ~25% is calculated from the $R$ vs. $H$ plots. The average RT switching current ($\langle I_c \rangle$) is 0.75 mA at a pulse width of 30 ms and the RT current density ($J_c$) calculated from $\langle I_c \rangle$ is 4.3 MA/cm$^2$. RA is estimated as 15 Ω·μm$^2$. The normalized typical bias dependence of the TMR is plotted in Fig. 2 for two samples with the MgO and $AlO_x$ barriers. Both samples show similar bias voltage (~450-600 mV) at the half of zero-bias TMR value, whereas a larger asymmetry in the bias dependence of the $AlO_x$ barrier is observed.

Since spin transfer induced magnetization switching is a thermally activated process [7-9], the switching current depends strongly on the pulse width used in measurements and the thermal factor ($K_u V / k_B T$) of the samples, where $K_u$ is the uniaxial anisotropy energy and $V$ is the volume of the free layer, both of which are dependent on the cell dimensions, $k_B$ is the Boltzmann constant and $T$ is the temperature of the sample. The value of the intrinsic switching current density $J_{c0}$ can be obtained by extrapolating the switching current dependence on the pulse width to 1 ns pulse width as described in details elsewhere [3, 7]. Figure 3 shows plots of the intrinsic switching current density $J_{c0}$ versus $TMR^{-1/2}$ for a set of MgO and $AlO_x$ MTJ nanostructures with TMR ranging from 15 to 150%, indicating systematic reduction of the intrinsic switching current density with increasing TMR. As can be seen in Fig. 3, the significant increase in TMR contributes to a reduction in $J_{c0}$ by almost a factor of 3. The average intrinsic switching current density $J_{c0}$ was found to be 2 – 3 MA/cm$^2$ for the junction structures with the MgO barriers and 5 – 12 MA/cm$^2$ for those with the $AlO_x$ barriers. Assuming a single-domain model of free layer with Slonczewski form of spin transfer torque [10, 11], the intrinsic switching current density $J_{c0}$ at zero temperature is given in terms of magnetic properties of a free layer:

$$J_{c0} = \frac{2e\alpha M_S t_F \left(H + H_K + 2\pi M_S\right)}{\hbar \eta} \tag{1}$$

where $e$ is the electron charge, $\alpha$ is the damping constant, $t_F$ is thickness of the free layer, $\hbar$ is the reduced Planck's constant, $\eta$ is spin transfer efficiency related to the TSP of incident spin polarized current by $\eta=(p/2)/(1+p^2 cos\theta)$ [11]. $H_k$ is the uniaxial anisotropy field of the free layer and $p$ represents the TSP from a reference (source) ferromagnetic layer and is considered as a constant [11]. A first-order approximation of the TSP can be found by taking $p_0$= $[\mathrm{TMR}/(2+\mathrm{TMR})]^{1/2}$, as derived at zero bias from Eq. (2) in Ref. 11 by assuming equal spin polarization on both sides of the barrier since identical electrode material (CoFeB) was used in this experiment. The intrinsic switching current $J_{c0}$ can then be estimated using Eq. (1) with $M_s$ = 800 emu/cm$^3$, $\alpha$ = 0.003 and $t_F$=2.2 nm. The calculated values of $J_{c0}^{P\to AP}$ and $J_{c0}^{AP\to P}$ as a function of $\mathrm{TMR}^{-1/2}$ are shown in Fig. 3 by lines for a set of MTJ samples with MgO and $AlO_x$ barriers with TMR ranging from 15 to 150%. Both experimental and calculated results indicate that the intrinsic switching current densities do decrease with increasing TMR and that in the low TMR limit $J_{c0}$ varies linearly with $\mathrm{TMR}^{-1/2}$. However, a noticeable discrepancy between the experiment and theory prediction exists, with implication that the TSP (obtained here from zero-bias TMR value) responsible for the spin transfer torque at finite bias might be overestimated since the TSP out of the reference (source) electrode is actually not separable from that of the free layer electrode (detector) and may be bias dependent. From the intrinsic current density ratio $J_{c0}^{P\to AP} / J_{c0}^{AP\to P}$, the effective spin polarization factor for spin transfer torque $p_{ST}$ is deduced and the data are plotted in Fig. 4(a), where the solid line represents the theoretical prediction [11] obtained by fitting the experimental data. The inset of Fig. 4(a) shows the correlation between $p_0$ and $p_{ST}$. The effective tunnel spin polarization, $p_{ST}$, is approximately two thirds of the polarization factor obtained from zero-bias TMR. This result provides actual effective TSP involved in spin transfer torque at finite bias. The upper bound of $p_{ST}$ is evaluated to be 0.22 for the MTJs with the $AlO_x$ barriers and 0.46 for those with the MgO barriers. In Fig. 4(b), the intrinsic switching current density is plotted as a function of the inverse of the spin transfer efficiency showing that the $J_{c0}^{P\rightarrow AP}$ or $J_{c0}^{AP\rightarrow P}$ varies almost linearly with $1/\eta$ in agreement with eq. (1). Note that for $J_{c0}^{P\rightarrow AP}$ estimate the data are limited for $\eta^{-1} > 4$, since $\eta \rightarrow$ ¼ as $p_{ST} \rightarrow 1$ in the extreme case. The shift of the data set reflects the asymmetry feature of the $J_{c0}^{P\rightarrow AP}$ and $J_{c0}^{AP\rightarrow P}$ [11].

The major challenge to fully understand the effective TSP at finite bias and its relationship to the spin transfer torque effect lies in the factorization of the conduction asymmetry [11] into the TSP at both sides of a barrier and their bias dependence. First, the TSP and its determination are not straightforward and cannot be derived from the TMR data since the TSP of the reference (source) electrode is inseparable from that of the free layer electrode (detector). Second, the bias dependence of the TSP in spin transfer torque remains unknown for these structures of interest. Interestingly, however, a recent spin polarized tunneling experiment [12] has showed a decreasing spin polarization factor for the electrons out of a reference ferromagnet CoFe with increasing bias. The spin polarization factor was decreased by up to one half upon increasing bias to 1 V, qualitatively following the essential features of the results of this study. The difference in the TSP as determined from TMR at zero bias and from the spin transfer magnetization switching measurements has implication that the $p_{ST}$ has a significant bias dependence. As a consequence of energy distribution of the injected electrons around the Fermi level, the band structure effect of the reference electrode combined with the interface between the electrode and a barrier may result in a bias dependence of the TSP. This is, however, not supported by the STM experiment results[13]. The observed decrease in the $p_{ST}$ at finite bias may be attributed to available electron trap states (either at finite bias or higher temperature in a dynamic bias-driven process) at the interfaces or within the insulating barriers that cause indirect tunneling [14]. Since these thermally or electrical stress generated states are not spin polarized and the indirect tunneling is spin-independent, the TSP for the electrons that tunnel through the barriers would become much smaller than expected. Further experiment is needed to address the details of this issue.

In summary, we have achieved spin-transfer driven magnetization switching at room temperature with an intrinsic switching current density $J_{c0}$ as low as 2-3 MA/cm$^2$ on MgO based MTJs with a TMR value as high as 150%. The $J_{c0}$ reduction found in MgO MTJs was about three times as compared to that obtained on $AlO_x$ MTJs, resulting from the enhancement of spin transfer efficiency caused by higher tunnel spin polarization. These low RA MgO MTJs enables high-density spin transfer switching MRAM (SpRAM) with fast access time of few nanoseconds. Moreover, we have experimentally determined TSP, ranging from 0.22 for $AlO_x$ to 0.46 for MgO based MTJs, responsible for the current-driven magnetization switching, and found that it is different from what one might expect theoretically. A brief of possible reasoning has been provided to stimulate more interest and future theoretical work in this field.

We would like to thank Wolfram Mass, Berthold Ocker and Juergen Langer of Singulus Inc. for joint development of low RA MgO and $AlO_x$ based MTJ films.

Figure Captions

Fig. 1. Typical RT field (a) and current (b) driven magnetization switching for an MTJ sample with an MgO barrier, and field (c) and current (d) driven magnetization switching for an MTJ sample with an $AlO_x$ barrier. Current pulse width of 30ms was used in obtaining (b) and (d)

Fig. 2. Typical normalized TMR bias dependence for the two MTJ samples with an MgO and an $AlO_x$ barrier, respectively.

Fig. 3. Intrinsic switching current density versus $TMR^{-1/2}$ for a set of MTJ samples with the MgO and $AlO_x$ barriers, in which open triangle (square) represents for $J_{c0}^{P\rightarrow AP}$ ( $J_{c0}^{AP\rightarrow P}$ ) in the MgO samples; and solid triangle (square) for $J_{c0}^{P\rightarrow AP}$ ( $J_{c0}^{AP\rightarrow P}$ ) in the $AlO_x$ samples.

Fig. 4. (a) Intrinsic switching current density ratio versus the TSP for a set of MTJ samples with the MgO and $AlO_x$ barriers, in which open (solid) square represents the data of the MgO ($AlO_x$) samples. In the inset, $p_{ST}$ is related to $p_0$. (b) Intrinsic switching current density versus spin polarization efficiency for a set of MTJ samples with the MgO and $AlO_x$ barriers, in which open triangle (square) represents for $J_{c0}^{P\rightarrow AP}$ ( $J_{c0}^{AP\rightarrow P}$ ) in the MgO samples; and solid triangle (square) for $J_{c0}^{P\rightarrow AP}$ ( $J_{c0}^{AP\rightarrow P}$ ) in the $AlO_x$ samples.

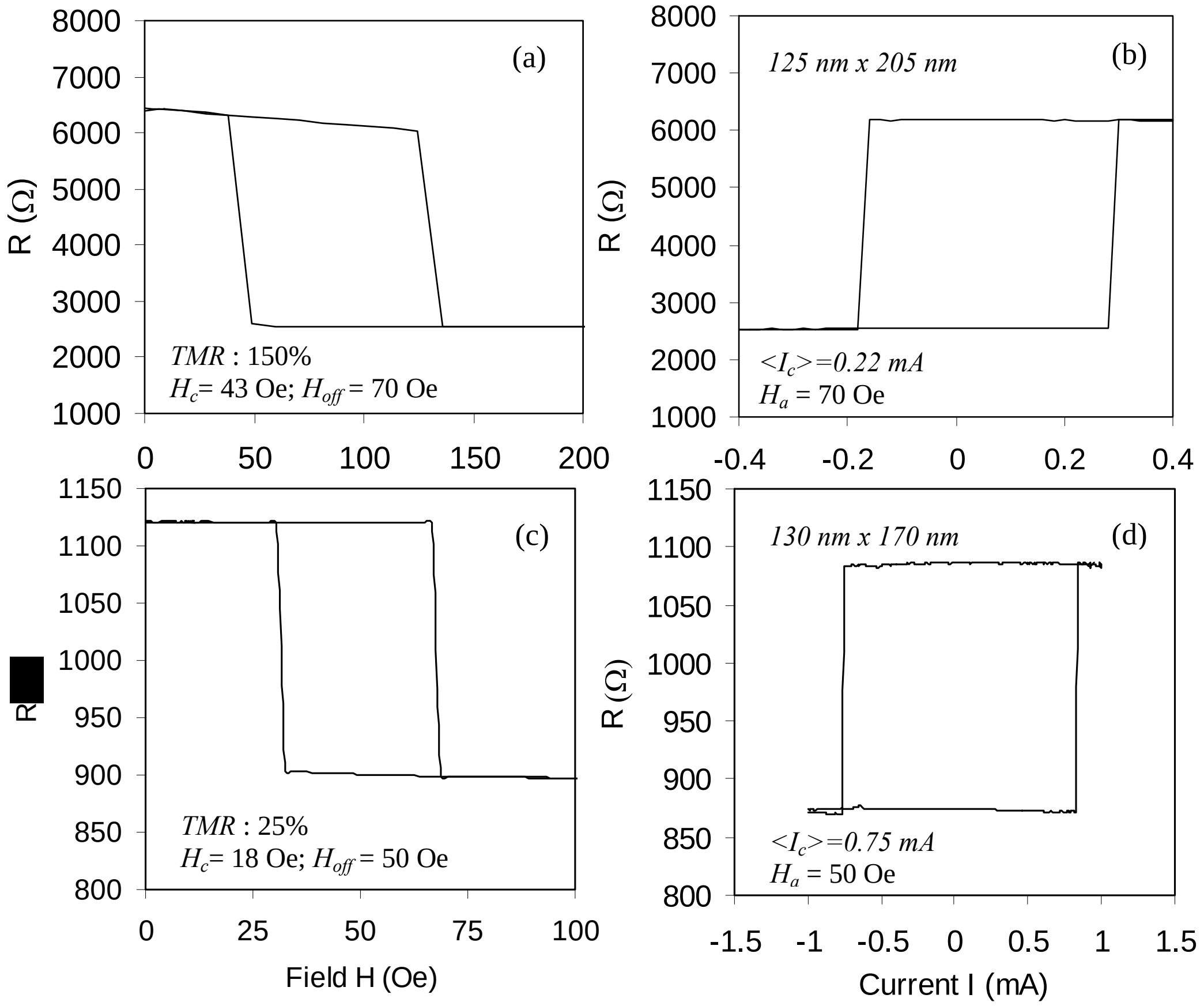

Fig. 1. Diao et.al

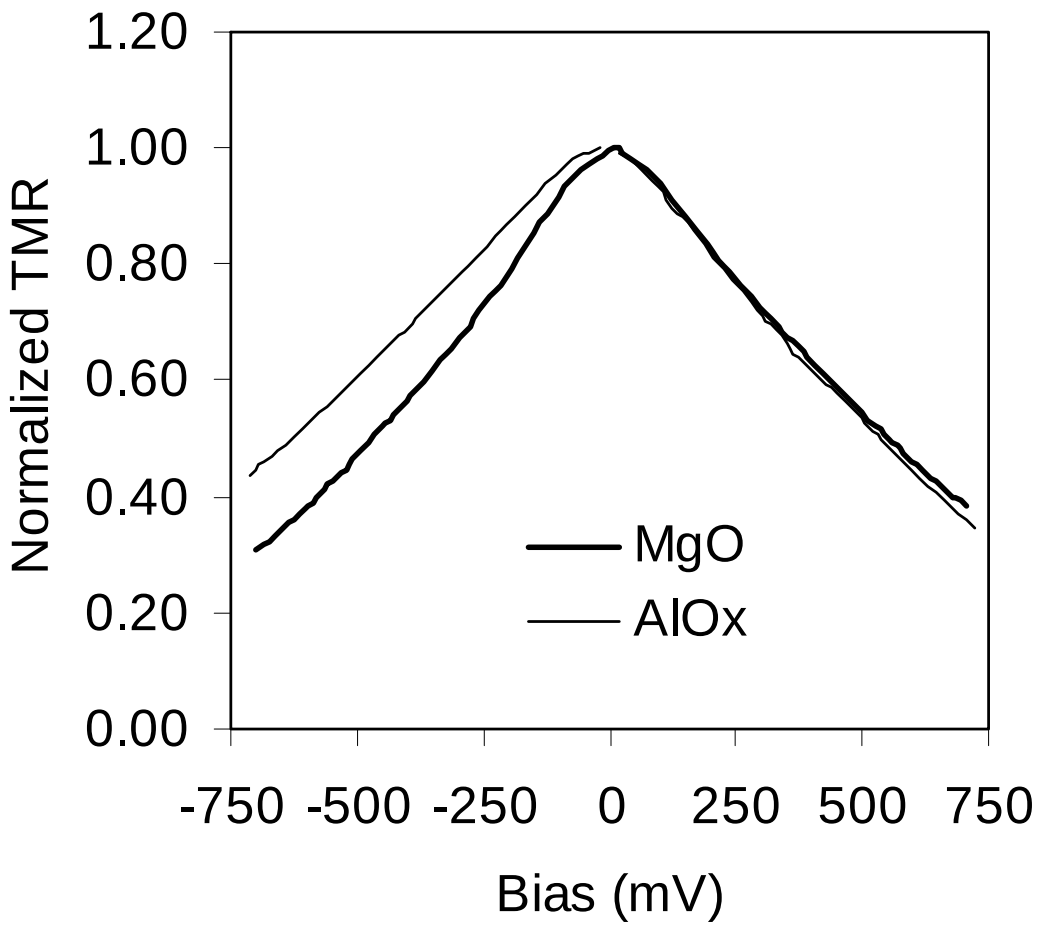

Fig. 2. Diao et.al

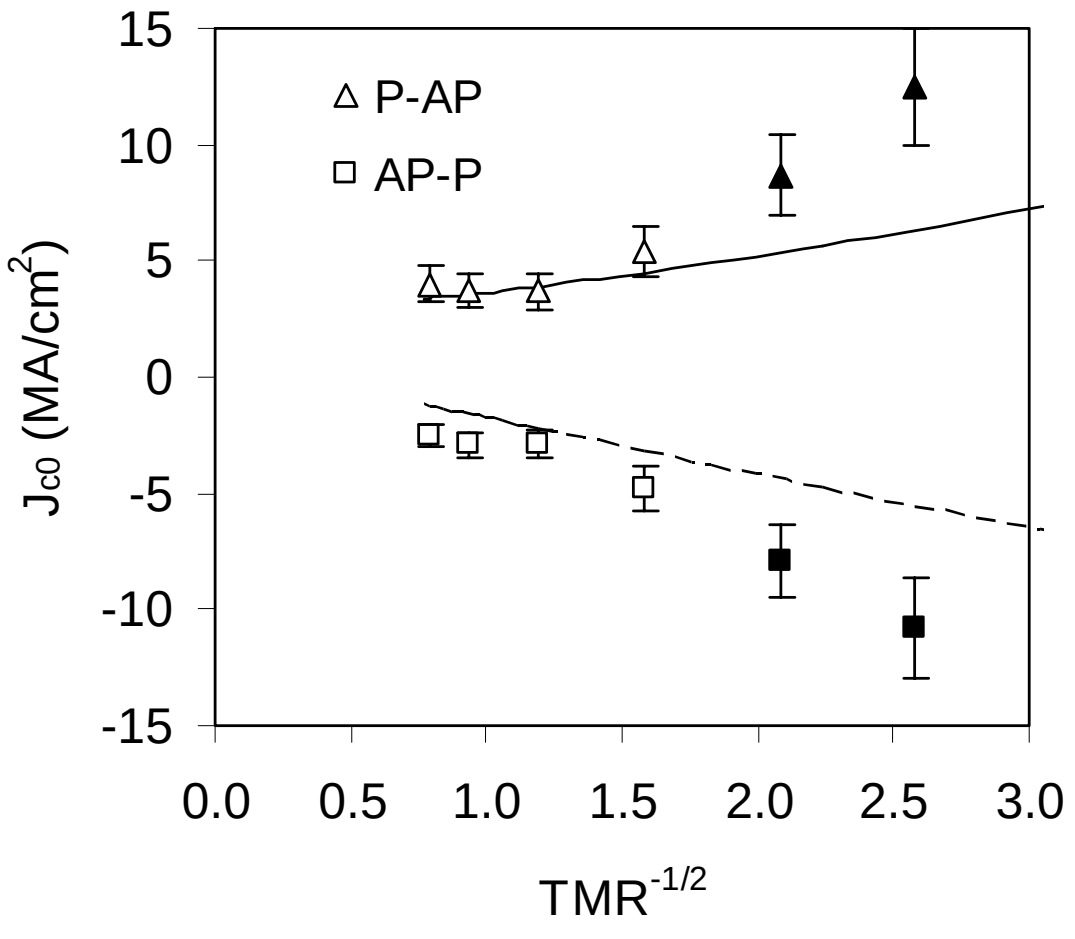

Fig. 3. Diao et.al

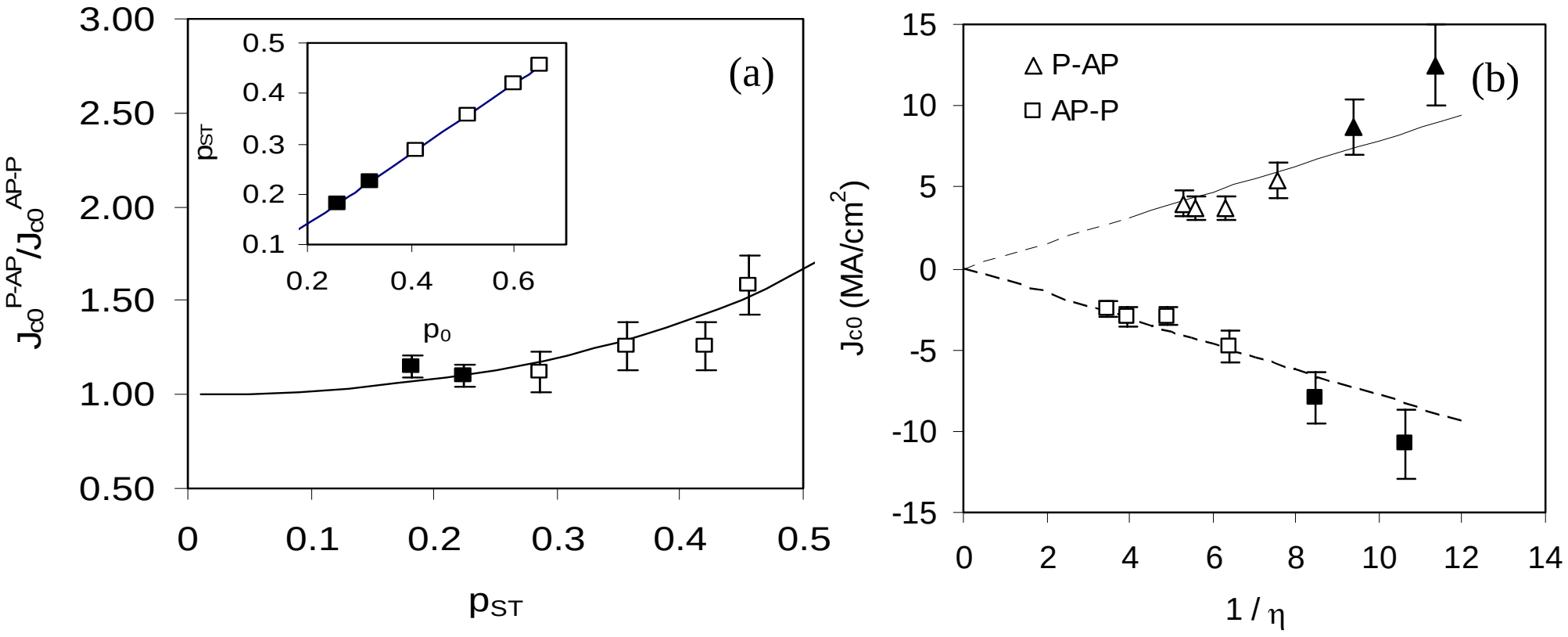

Fig. 4. Diao et.al